\pdfoutput=1
\documentclass[11pt,a4paper]{article}

\usepackage[utf8]{inputenc}
\usepackage[T1]{fontenc}
\usepackage[english]{babel}

\usepackage[
  left=25mm, right=25mm,
  top=30mm, bottom=30mm
]{geometry}

\usepackage{mathpazo}          
\usepackage{inconsolata}       
\usepackage{amsmath,amssymb}

\usepackage{graphicx}
\usepackage{booktabs}
\usepackage{multirow}
\usepackage{caption}
\usepackage{subcaption}

\usepackage[ruled,vlined,linesnumbered]{algorithm2e}
\SetKwInput{KwIn}{Input}
\SetKwInput{KwOut}{Output}

\usepackage[colorlinks,
  linkcolor=blue!60!black,
  citecolor=blue!60!black,
  urlcolor=blue!60!black
]{hyperref}
\usepackage{url}

\usepackage{tikz}
\usetikzlibrary{
  arrows.meta,
  positioning,
  fit,
  backgrounds,
  calc,
  shapes.geometric,
}

\usepackage{enumitem}
\usepackage{xcolor}
\usepackage{microtype}

\usepackage{natbib}
\title{%
  MemX: A Local-First Long-Term Memory System\\
  for AI Assistants%
}

\author{
  Lizheng Sun
}

\date{March 2026}

\begin{document}
\maketitle

\begin{abstract}
We present \textbf{MemX}, a local-first long-term memory system
for AI assistants with stability-oriented retrieval design.
MemX is implemented in Rust on top of libSQL and an
OpenAI-compatible embedding API, providing persistent,
searchable, and explainable memory for conversational agents.
Its retrieval pipeline applies vector recall, keyword recall,
Reciprocal Rank Fusion (RRF), four-factor re-ranking, and a
low-confidence rejection rule that suppresses spurious recalls
when no answer exists in the memory store.

We evaluate MemX on two axes.
First, two custom Chinese-language benchmark suites (43 queries,
${\leq}1{,}014$ records) validate pipeline design:
$\text{Hit@1}{=}91.3\%$ on a default scenario and $100\%$
under high confusion, with conservative miss-query suppression.
Second, the LongMemEval benchmark (500 queries, up to
220{,}349 records) quantifies system boundaries across four
ability types and three storage granularities.
At fact-level granularity the system reaches
$\text{Hit@5}{=}51.6\%$ and $\text{MRR}{=}0.380$, doubling
session-level performance, while temporal and multi-session
reasoning remain challenging (${\leq}43.6\%$ Hit@5).
FTS5 full-text indexing reduces keyword search latency by
$1{,}100{\times}$ at 100k-record scale, keeping end-to-end
search under 90\,ms.

Unlike Mem0 and related work that targets end-to-end agent
benchmarks, MemX focuses on a narrower, reproducible baseline:
local-first deployment, structural simplicity, explainable
retrieval, and stability-oriented design.

\end{abstract}

\section{Introduction}
\label{sec:intro}

Large language models have achieved strong single-turn comprehension,
yet they remain stateless across sessions.
Without a persistent memory layer, an AI assistant cannot reliably
retain user preferences, project conventions, operational procedures,
incident resolutions, or domain-specific constraints.
This gap is especially pronounced for \emph{local} AI assistants,
where users expect the system to accumulate knowledge over time,
avoid repetitive questioning, and refrain from fabricating
answers when no relevant memory exists.
The challenge of equipping LLMs with long-term memory has been
widely recognized as a key research
direction~\citep{zhang2024survey}.
While retrieval-augmented generation
(RAG)~\citep{lewis2020rag} has become the dominant paradigm for
grounding LLM outputs in external knowledge, most RAG systems
target document corpora rather than the incremental, personalized
memories that accumulate during daily assistant interactions.

Recent work has begun to address this gap through diverse
architectural choices.
Mem0 \citep{chhikara2025mem0} frames long-term memory as an
integral component of agent infrastructure, covering memory
extraction, update, retrieval, and downstream task performance.
MemGPT~\citep{packer2024memgpt} introduces an operating-system
metaphor for managing LLM memory hierarchies.
Generative Agents~\citep{park2023generative} demonstrate how
persistent memory enables believable social simulation.
MemLLM~\citep{modarressi2024memllm} takes a complementary
fine-tuning approach, augmenting the model itself with an
explicit read--write memory module.
A recent survey by \citet{gao2024rag_survey} highlights the
growing diversity of RAG architectures, yet notes that most
systems assume access to a pre-existing knowledge base
rather than incrementally constructed personal memories.

Effective memory retrieval requires combining semantic and
lexical signals.
Dense vector search, typically organized through approximate
nearest-neighbor structures such as
HNSW \citep{malkov2020hnsw}, captures meaning-level
similarity but can miss keyword-specific cues.
Sparse lexical methods, from classical
BM25~\citep{lin2021bm25} to lightweight full-text indexes
such as SQLite FTS5, complement dense retrieval by matching
exact terms.
Recent embedding models such as
BGE-M3~\citep{chen2024bge_m3} natively support both dense
and sparse representations, underscoring the importance of
hybrid retrieval.
Reciprocal Rank Fusion (RRF)~\citep{cormack2009rrf} provides
a simple, parameter-light method for combining ranked lists
from heterogeneous retrievers, making it well-suited to
systems that pair vector and full-text indexes.

Most existing memory-augmented systems assume cloud-hosted
deployment with ample compute and centralized storage, and
their evaluations emphasize end-to-end agent task completion
rather than retrieval quality in isolation.
However, the local-first software
paradigm~\citep{kleppmann2019localfirst} argues that users
should retain full ownership of their data, with the ability
to operate offline and without depending on third-party services.
For a personal AI assistant, local-first deployment offers
concrete advantages: lower latency, stronger privacy guarantees,
and independence from external infrastructure changes.
Yet a narrower but critical gap remains: under these constraints,
how should a memory system balance recall against the risk of
spurious results when no relevant memory exists?

This gap motivates the following research question:

\begin{quote}
\emph{In a single-user, locally-deployed AI assistant scenario,
can a structurally simple, explainable memory system
achieve stable recall of relevant memories while
minimizing spurious recalls on unanswerable queries?}
\end{quote}

We answer this question with MemX, a system built on four design
principles (local-first deployment, structural simplicity,
real-embedding evaluation, and stability-over-recall) and
validate it on two custom benchmark suites with live embedding
APIs as well as the external LongMemEval
benchmark~\citep{wang2024longmemeval}
(500 queries, up to 220k records).

\paragraph{Contributions.}
\begin{enumerate}[leftmargin=*,nosep]
  \item A complete implementation of a local long-term memory
        system featuring hybrid retrieval with explicit
        access/retrieval tracking separation.
  \item A reproducible benchmark framework that directly invokes
        internal search functions with real embeddings, supporting
        scenario extension, threshold sweeping, and structured
        JSON reporting.
  \item Empirical evidence on two custom scenarios (43 queries)
        characterizing the pipeline's strengths (stable single-topic
        recall, conservative miss handling) and boundaries
        (multi-topic coverage gaps, high-confusion false recalls).
  \item A storage-granularity study on
        LongMemEval~\citep{wang2024longmemeval} (500 queries,
        up to 220k records) showing that fact-level chunking
        doubles retrieval quality over session-level storage,
        and quantifying per-ability-type performance boundaries.
\end{enumerate}

\section{Related Work}
\label{sec:related}

\paragraph{Long-term memory for LLM agents.}
The need for persistent memory in LLM-based systems has been
widely recognized~\citep{zhang2024survey}.
Mem0~\citep{chhikara2025mem0} proposes a production-oriented
memory layer with extraction, update, and retrieval stages,
evaluated on end-to-end agent tasks.
MemGPT~\citep{packer2024memgpt} draws an analogy to virtual memory,
allowing an LLM to page information in and out of its context window.
MemLLM~\citep{modarressi2024memllm} fine-tunes models to use
an explicit read-write memory.
MemX differs from these systems in scope: rather than
targeting general agent benchmarks, we focus on a
local, single-user deployment with reproducible retrieval evaluation.

\paragraph{Retrieval-augmented generation.}
RAG~\citep{lewis2020rag} established the paradigm of
conditioning generation on retrieved passages.
Subsequent surveys~\citep{gao2024rag_survey} document the rapid
diversification of RAG architectures.
MemX can be viewed as a specialized RAG store whose retrieval
pipeline is optimized for stability rather than maximum recall.

\paragraph{Vector and hybrid retrieval.}
FAISS~\citep{johnson2019faiss} and
HNSW~\citep{malkov2020hnsw} are foundational approximate
nearest-neighbor methods.
Hybrid approaches that combine dense and sparse
retrieval—fused via Reciprocal Rank
Fusion~\citep{cormack2009rrf}—have shown consistent benefits
in information retrieval~\citep{lin2021bm25,chen2024bge_m3}.
MemX adopts RRF as the fusion step and adds a four-factor
re-ranking stage tailored to memory retrieval.

\paragraph{Graph-enhanced memory.}
GraphRAG~\citep{edge2024graphrag} and related
efforts~\citep{guo2024knowledgegraph_survey} enrich retrieval
with graph structure.
Mem0's later work also explores graph-augmented memory where
contradiction, supersession, and causality links participate
in retrieval.
MemX maintains a link table (\texttt{memory\_links}) supporting
seven relation types, but does not yet integrate multi-hop
traversal or graph-aware re-ranking into the search pipeline;
we note this as a planned extension
(Section~\ref{sec:discussion}).

\paragraph{Local-first software.}
\citet{kleppmann2019localfirst} articulate the principles of
local-first software: data ownership, offline capability, and
low-latency access.
MemX embodies these principles by storing all data in a single
libSQL~\citep{libsql2023} file on the user's machine.

\section{System Design}
\label{sec:system}

\subsection{Architecture Overview}
\label{sec:arch}

\begin{figure}[t!]
\centering
\includegraphics[width=0.95\linewidth]{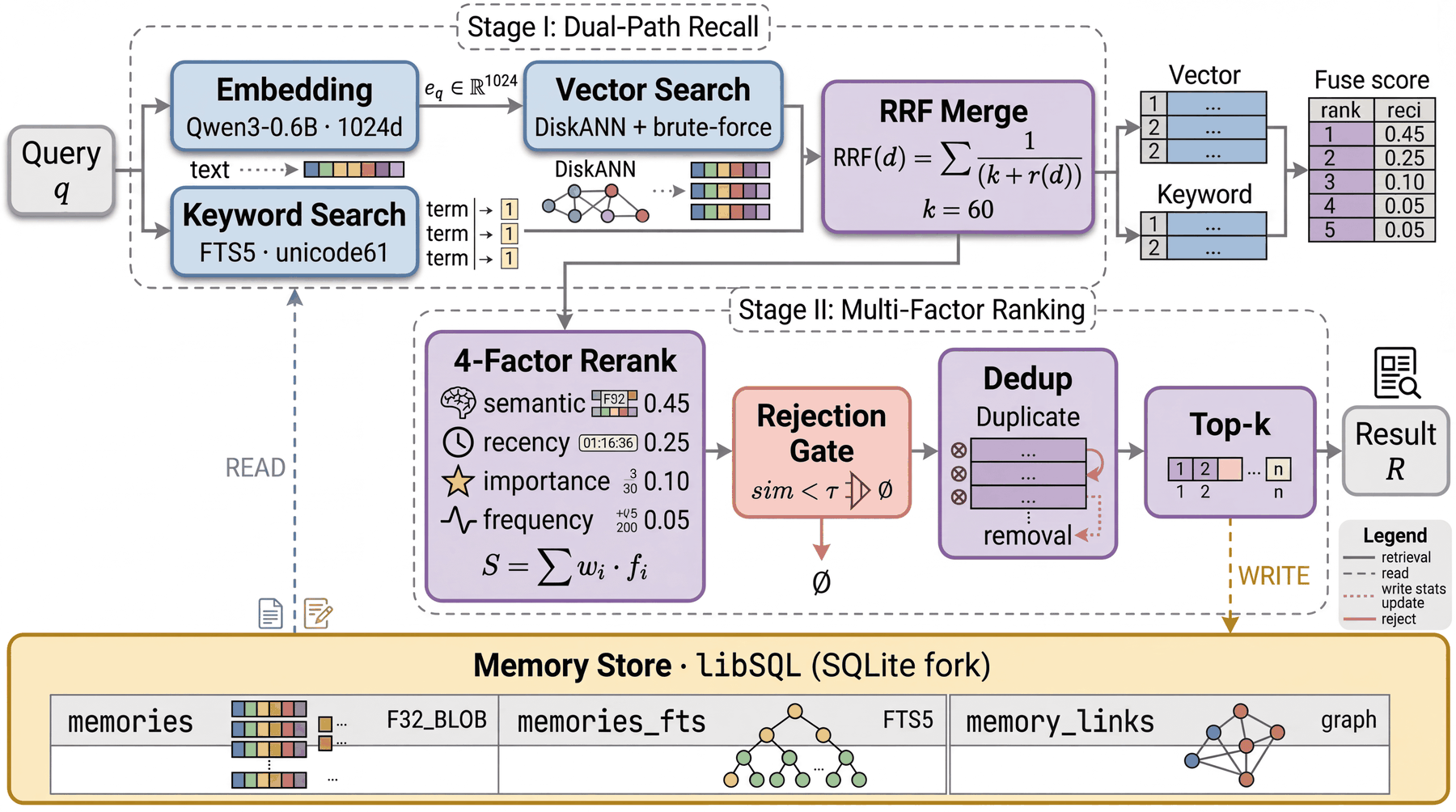}
\caption{%
  MemX search pipeline.
  A query is embedded by Qwen3-0.6B (1024-dim) and routed to
  two parallel recall paths: DiskANN/brute-force vector search
  and FTS5 keyword matching.
  Results are fused via RRF ($k{=}60$), re-ranked by four
  weighted factors (semantic similarity, recency, importance,
  frequency), and filtered by a rejection gate that returns
  $\varnothing$ when similarity falls below threshold~$\tau$.
  After deduplication the top-$k$ results are returned and
  retrieval statistics are written back to the database.%
}
\label{fig:architecture}
\end{figure}

Figure~\ref{fig:architecture} illustrates the end-to-end
search pipeline.
A query enters the system and is simultaneously processed by
a vector recall path (via an OpenAI-compatible embedding API)
and a keyword recall path.
The two candidate sets are merged using RRF, re-ranked by a
four-factor scoring function, filtered by a low-confidence
rejection rule, and finally the top-$k$ results are returned
with retrieval statistics recorded.

The system is implemented in Rust.
Storage uses libSQL~\citep{libsql2023}, an open fork of
SQLite~\citep{hipp2000sqlite}, which provides a single-file
database with vector-search extensions.

\subsection{Data Model}
\label{sec:datamodel}

The primary table \texttt{memories} stores each memory entry
with content, embedding, type, tags, metadata, and importance
fields, together with access and retrieval counters
and their corresponding timestamps.

A secondary table \texttt{memory\_links} records directed
relations between memories.
Seven link types are currently defined:
\textit{similar}, \textit{related}, \textit{contradicts},
\textit{extends}, \textit{supersedes}, \textit{caused\_by},
and \textit{temporal}.
Although the graph infrastructure exists, multi-hop traversal
and graph-aware retrieval are not yet integrated into the
search pipeline.

\subsection{Hybrid Search Pipeline}
\label{sec:pipeline}

The search pipeline is deterministic and proceeds in the
fixed stages formalized in Algorithm~\ref{alg:search}.

\begin{algorithm}[t]
\caption{Hybrid Search with Low-Confidence Rejection}
\label{alg:search}
\KwIn{Query $q$; candidate limit $n$; rejection threshold $\tau$; result limit $k$}
\KwOut{Ranked memory set $\mathcal{R}$ (possibly empty)}
$\mathbf{e}_q \gets \textsc{Embed}(q)$\;
$\mathcal{V} \gets \textsc{VectorRecall}(\mathbf{e}_q,\; n)$ \tcp*{cosine similarity}
$\mathcal{K} \gets \textsc{KeywordRecall}(q,\; n)$ \tcp*{FTS5 full-text match}
$\mathcal{C} \gets \textsc{RRF}(\mathcal{V},\; \mathcal{K})$
  \tcp*{Reciprocal Rank Fusion}
$\mathcal{S} \gets \textsc{FourFactorRerank}(\mathcal{C})$
  \tcp*{Eq.\,\ref{eq:rerank}--\ref{eq:frequency}}
$\hat{\mathcal{S}} \gets \textsc{ZScoreNormalize}(\mathcal{S})$
  \tcp*{Eq.\,\ref{eq:zscore}}
\If{$\mathcal{K} = \varnothing$ \textbf{and}
    $\max_{m \in \mathcal{V}} \mathrm{sim}(\mathbf{e}_q, \mathbf{e}_m) < \tau$}{
  \Return{$\varnothing$} \tcp*{reject low-confidence result}
}
$\mathcal{D} \gets \textsc{Dedup}(\hat{\mathcal{S}})$
  \tcp*{content + tag-signature}
$\mathcal{R} \gets \textsc{TopK}(\mathcal{D},\; k)$\;
$\textsc{RecordRetrievalStats}(\mathcal{R})$\;
\Return{$\mathcal{R}$}\;
\end{algorithm}

\subsection{Four-Factor Re-ranking}
\label{sec:rerank}

After RRF fusion, each candidate memory $m$ is assigned a
composite score:
\begin{equation}
  \label{eq:rerank}
  \mathrm{score}(m) =
    \alpha_s \cdot f_{\text{sem}}(m)
  + \alpha_r \cdot f_{\text{rec}}(m)
  + \alpha_f \cdot f_{\text{freq}}(m)
  + \alpha_i \cdot f_{\text{imp}}(m)
\end{equation}
where the four factors and their default weights are:

\begin{table}[ht]
\centering
\small
\caption{Default re-ranking weights.}
\label{tab:weights}
\begin{tabular}{@{}llr@{}}
\toprule
\textbf{Factor} & \textbf{Signal source} & \textbf{Weight} \\
\midrule
$f_{\text{sem}}$  & Semantic similarity (from RRF score) & $\alpha_s = 0.45$ \\
$f_{\text{rec}}$  & Recency (prefer \texttt{last\_retrieved\_at}, & $\alpha_r = 0.25$ \\
                   & fallback to \texttt{created\_at}) & \\
$f_{\text{freq}}$ & Frequency (prefer \texttt{retrieval\_count}, & $\alpha_f = 0.05$ \\
                   & fallback to \texttt{access\_count}) & \\
$f_{\text{imp}}$  & Importance (explicit annotation) & $\alpha_i = 0.10$ \\
\bottomrule
\end{tabular}
\end{table}

\noindent
The weights sum to $0.85$, not $1.0$; because the subsequent
z-score normalization (Equation~\ref{eq:zscore}) removes
absolute scale dependence, only the \emph{ratios} among weights
affect the final ranking.  The remaining capacity leaves room
for additional factors in future extensions.

The recency factor uses \texttt{last\_retrieved\_at} rather than
\texttt{last\_accessed\_at}, and the frequency factor uses
\texttt{retrieval\_count} rather than \texttt{access\_count}.
Ranking therefore reflects how often a memory has been
\emph{useful in search} rather than how often it has been
\emph{explicitly viewed}, avoiding signal pollution from
administrative access.

\paragraph{Factor computation.}
Each factor is computed from the memory's metadata.
Let $t_m$ be the effective timestamp
(prefer \texttt{last\_retrieved\_at}, fall back to
\texttt{created\_at}) and $c_m$ the effective count
(prefer \texttt{retrieval\_count}, fall back to
\texttt{access\_count}).
Recency follows an exponential half-life decay:
\begin{equation}
  \label{eq:recency}
  f_{\text{rec}}(m) = 2^{-\,d_m / h}
\end{equation}
where $d_m = (\mathit{now} - t_m) / 86400$ is the age in days and
$h$ is a configurable half-life (default $h = 30$ days).
This gives a smooth, parameter-transparent curve: a memory's
recency score halves every $h$ days with no hard cutoffs.
Frequency is log-normalized and capped:
\begin{equation}
  \label{eq:frequency}
  f_{\text{freq}}(m) = \min\!\bigl(1,\;
    \ln(c_m + 1)\,/\,10\bigr)
\end{equation}
The logarithm prevents highly-accessed memories from dominating
the score, while the cap ensures the factor stays in $[0,1]$.

\paragraph{Score normalization.}
After the composite scores are computed across all candidates,
the system applies z-score normalization followed by a sigmoid
transformation:
\begin{equation}
  \label{eq:zscore}
  \hat{s}(m) = \operatorname{sigm}\!\Bigl(
    \frac{s(m) - \mu}{\sigma_s}
  \Bigr)
  = \frac{1}{1 + \exp\!\bigl(
    -(s(m) - \mu)\,/\,\sigma_s
  \bigr)}
\end{equation}
where $\mu$ and $\sigma_s$ are the mean and standard deviation
of $\{s(m)\}$ over the candidate set.
Z-score centering makes scores comparable across queries with
different candidate-set distributions; the subsequent sigmoid
compresses outliers while spreading scores near the center,
improving the discriminability of the final ranking.
If $\sigma_s < 10^{-6}$ (all candidates score identically),
normalization is skipped to avoid division instability.

\subsection{Result Deduplication}
\label{sec:dedup}

Template-based memory stores and repeated ingestion can produce
near-identical records that, without intervention, would occupy
multiple top-$k$ slots and reduce topic diversity.
The system applies two deduplication layers after scoring:

\begin{enumerate}[nosep]
  \item \textbf{Content deduplication.}
    Memories with identical trimmed content are collapsed;
    only the highest-scoring instance is retained.
  \item \textbf{Tag-signature deduplication.}
    A \emph{tag signature} is defined as the concatenation of a
    memory's type and its sorted tag set
    (e.g., \texttt{procedural::ops|release}).
    At most one memory per tag signature is returned.
    This prevents a cluster of same-topic records (common when
    procedural memories share identical tags) from crowding out
    results from other topics.
\end{enumerate}
Memories without tags are exempt from tag-signature filtering
and compete solely on score.
In the benchmark's $3{\times}$-scaled datasets (up to 1{,}014
records), this mechanism matters in practice:
without it, the top-5 results for procedural queries consist
entirely of template variants of the same topic,
and Coverage@5 drops accordingly.

\subsection{Low-Confidence Rejection Rule}
\label{sec:rejection}

To suppress spurious recalls when the memory store contains no
relevant entry, the system applies a conservative rejection rule:
\begin{quote}
  If the keyword recall set $\mathcal{K}$ is empty \textbf{and}
  the maximum vector similarity score is below threshold $\tau$,
  return an empty result set.
\end{quote}
The current recommended threshold is $\tau = 0.50$, determined
empirically through the threshold sweep experiment described in
Section~\ref{sec:sweep}.
This rule trades a small amount of recall for a meaningful
reduction in false positives on unanswerable queries.
Section~\ref{sec:rejection_space} formalizes five candidate
rejection rules and evaluates them against the benchmark data.

\subsection{Access vs.\ Retrieval Separation}
\label{sec:tracking}

The system explicitly distinguishes two types of interactions:
\begin{description}[leftmargin=1em,nosep,style=sameline]
  \item[Access] — A user or system explicitly reads a
    memory entry (e.g., viewing details).
    Tracked by \texttt{access\_count} and
    \texttt{last\_accessed\_at}.
  \item[Retrieval] — A memory is returned as a search
    result.
    Tracked by \texttt{retrieval\_count} and
    \texttt{last\_retrieved\_at}.
\end{description}
This separation prevents explicit reads from inflating
retrieval-based ranking signals.
Section~\ref{sec:separation_impact} quantifies the ranking impact
of this design choice through a worked numerical example.

\section{Benchmark Framework}
\label{sec:benchmark}

\subsection{Design Rationale}
\label{sec:bench_rationale}

Many memory system evaluations rely on single demonstrations or
manually curated examples, which suffice for showcasing behavior
but are insufficient for locating failure modes.
We implement a standalone benchmark framework with the following
properties:
\begin{itemize}[nosep]
  \item Directly invokes the system's internal search functions
        (bypassing HTTP routing), measuring intrinsic retrieval
        performance.
  \item Uses a live OpenAI-compatible embedding API rather than
        pre-computed or random vectors.
  \item Automatically constructs an isolated test database per run.
  \item Outputs structured JSON reports supporting downstream
        analysis.
  \item Supports threshold sweeping and scenario extension.
\end{itemize}

\subsection{Scenario Design}
\label{sec:scenarios}

We construct two benchmark scenarios in Chinese, reflecting
the target deployment language.

\paragraph{Default scenario.}
This scenario models realistic usage of a local AI assistant,
covering user preferences (e.g., coffee habits), job role,
code style conventions, deployment checklists, client demo
constraints, database incident procedures, Rust debugging
strategies, budget thresholds, travel preferences, and
multiple noise topics.
After $3{\times}$ scaling, the scenario contains 1{,}014 memory
records and 25 queries spanning seven query kinds:
\textit{keyword\_exact}, \textit{semantic\_paraphrase},
\textit{procedure\_recall}, \textit{long\_context},
\textit{reflective}, \textit{multi\_fact}, and \textit{miss}.

\paragraph{High-confusion scenario.}
This scenario deliberately places semantically overlapping
topics adjacent to each other:
local deployment vs.\ vendor switching messaging,
migration root cause vs.\ release compatibility checks,
search stability vs.\ search speed preferences,
budget caps vs.\ ROI exceptions, and
Rust borrow errors vs.\ trait bound errors.
After $3{\times}$ scaling it contains 600 records and 18 queries.
It is designed to stress-test false-recall suppression,
multi-topic coverage, and miss-query convergence.

\paragraph{LongMemEval scenario.}
To evaluate at larger scale, we adopt the LongMemEval
benchmark~\citep{wang2024longmemeval}, which provides 500
annotated questions derived from 19{,}195 multi-turn
conversation sessions.
Questions span four ability types: \textit{information
extraction}, \textit{knowledge update},
\textit{multi-session reasoning}, and \textit{temporal reasoning}.
We convert the dataset into three MemX scenarios at different
storage granularities (session, round, fact); fact-level
extraction uses DeepSeek-Chat to decompose each session into
atomic statements.
Details and results are presented in
Section~\ref{sec:longmemeval}.

\paragraph{Total query budget.}
Across all scenarios the evaluation uses 543 queries:
43 from the two custom scenarios
(25 default $+$ 18 high-confusion) and 500 from LongMemEval.

\subsection{Metrics}
\label{sec:metrics}

The benchmark reports the following metrics
(Table~\ref{tab:metrics}):

\begin{table}[ht]
\centering
\small
\caption{Benchmark metrics.}
\label{tab:metrics}
\begin{tabular}{@{}lp{8.5cm}@{}}
\toprule
\textbf{Metric} & \textbf{Description} \\
\midrule
Hit@$k$ ($k{=}1,3,5$) &
  Fraction of relevant queries where at least one correct memory
  appears in the top-$k$ results. \\
Topic Coverage@$k$ &
  For multi-topic queries, the fraction of expected topics covered
  in the top-$k$ results. \\
MRR &
  Mean Reciprocal Rank over relevant queries. \\
Miss-Empty-Rate &
  Fraction of miss queries (no correct answer exists) for which
  the system returns an empty result set. \\
Miss-Strict-Rate &
  Fraction of miss queries for which the top vector score stays
  below the miss threshold (more stringent than Miss-Empty-Rate). \\
Avg/P95 Search (ms) &
  Average and 95th-percentile end-to-end search latency. \\
\bottomrule
\end{tabular}

\medskip\noindent
For binary metrics (Hit@$k$, Miss-Empty-Rate, Miss-Strict-Rate)
we report 95\% Wilson score confidence intervals where the
sample size is finite.  The Wilson interval for a proportion
$\hat{p}$ observed in $n$ trials is
\[
  \frac{
    \hat{p} + \frac{z^2}{2n}
    \pm z \sqrt{\frac{\hat{p}(1-\hat{p})}{n}
               + \frac{z^2}{4n^2}}
  }{1 + \frac{z^2}{n}}\,,
  \qquad z = 1.96\,.
\]
This interval has better coverage than the normal approximation
when $n$ is small or $\hat{p}$ is near 0 or 1~\citep{wilson1927}.
\end{table}

\section{Experiments}
\label{sec:experiments}

\subsection{Setup}
\label{sec:setup}

All experiments use the following configuration:
\begin{itemize}[nosep]
  \item \textbf{Embedding model:} Qwen3-Embedding-0.6B~\citep{qwen3embedding2025}
  \item \textbf{Embedding dimension:} 1{,}024
  \item \textbf{Rejection threshold:} $\tau = 0.50$
  \item \textbf{Re-ranking weights:} as in Table~\ref{tab:weights}
  \item \textbf{Vector index:} DiskANN (via libSQL vector extension)
  \item \textbf{Keyword index:} FTS5 with \texttt{unicode61} tokenizer
\end{itemize}
The benchmark framework invokes internal search functions
directly; reported latencies therefore reflect intrinsic
retrieval cost, not full HTTP round-trip time.

\subsection{Scenario-Level Results}
\label{sec:results}

Table~\ref{tab:main_results} summarizes the key metrics for
both scenarios.

\begin{table}[t]
\centering
\caption{Retrieval quality and latency across the two benchmark
scenarios ($\tau = 0.50$).
95\% Wilson confidence intervals are shown in brackets for
binary metrics.}
\label{tab:main_results}
\begin{tabular}{@{}lrr@{}}
\toprule
\textbf{Metric}
  & \textbf{Default}
  & \textbf{High-Confusion} \\
\midrule
Records             & 1{,}014            & 600              \\
Queries (rel / miss)& 23 / 2             & 14 / 4           \\
\midrule
Hit@1               & 91.3\% [73, 98]    & 100.0\% [78, 100] \\
Hit@5               & 95.7\% [79, 99]    & 100.0\% [78, 100] \\
Coverage@5          & 91.3\%             & 75.0\%            \\
MRR                 & 0.935              & 1.000             \\
\midrule
Miss-Empty-Rate     & 50.0\% [9, 91]     & 75.0\% [30, 95]   \\
Miss-Strict-Rate    & 100.0\% [34, 100]  & 75.0\% [30, 95]   \\
\midrule
Avg Search (ms)     & 538.32             & 397.68            \\
P95 Search (ms)     & 679.01             & 459.84            \\
\bottomrule
\end{tabular}
\end{table}

\paragraph{Default scenario.}
With the expanded 25-query set, Hit@1 reaches 91.3\%
(Wilson 95\% CI [73, 98]\%) and MRR is 0.935.
The two misses at rank~1 are keyword-exact queries whose
target memories share vocabulary with higher-scoring
distractors: \texttt{travel\_network} is recovered at rank~2
(Hit@3 = 95.7\%), while \texttt{deploy\_rollback} is not
recovered within the top-5.
Coverage@5 is 91.3\%, with gaps on the two
\textit{multi\_fact} queries that each require two distinct
topics.
Of the two miss queries, one (\textit{``What is the user's
favorite gym brand?''}) is correctly rejected
($v_{\max}{=}0.453$), while the other (\textit{``What breed
is the user's pet?''}) produces a borderline
$v_{\max}$ that exceeds $\tau$, giving
Miss-Empty-Rate = 50\% [9, 91]\%.
Miss-Strict-Rate remains 100\% [34, 100]\%.

\paragraph{High-confusion scenario.}
Hit@1 is 100\% [78, 100]\% across all 14 relevant queries,
confirming that the system reliably identifies the primary
relevant topic even under high semantic overlap.
Coverage@5 is 75.0\%, reflecting the difficulty of
multi-topic queries where each expected topic competes with
a semantically similar neighbor.
Of the four miss queries, three are correctly rejected
(Miss-Empty-Rate = 75\% [30, 95]\%), while
\textit{``Which public cloud region does the client
require?''} produces $v_{\max}{=}0.621$, exceeding $\tau$
and resulting in a spurious recall of the
\texttt{local\_storage\_policy} topic.
This failure case is analyzed further in
Section~\ref{sec:failures}.

\subsection{Threshold Sweep}
\label{sec:sweep}

We sweep the rejection threshold $\tau$ over
$\{0.48, 0.50, 0.52, 0.64\}$, running both scenarios at each
value.
Table~\ref{tab:sweep} reports the cross-scenario averages.

\begin{table}[t]
\centering
\caption{Threshold sweep results (averaged across both scenarios,
43 queries).}
\label{tab:sweep}
\begin{tabular}{@{}rrrrl@{}}
\toprule
$\tau$ & Avg Hit@1 & Avg Miss-Empty & Avg Miss-Strict
  & Note \\
\midrule
0.48 &  95.7\% & 62.5\% & 87.5\% & Tied with 0.50 \\
0.50 &  95.7\% & 62.5\% & 87.5\% & \textbf{Recommended} \\
0.52 &  93.5\% & 62.5\% & 87.5\% & Begins to hurt recall \\
0.64 &  87.0\% & 100.0\% & 87.5\% & Significantly hurts recall \\
\bottomrule
\end{tabular}
\end{table}

At $\tau = 0.48$ and $\tau = 0.50$, Avg~Hit@1 is highest at
95.7\% with identical Miss-Empty-Rate (62.5\%) and
Miss-Strict-Rate (87.5\%).
Raising the threshold to $\tau = 0.52$ causes one additional
false rejection (the \texttt{deploy\_rollback} query in the
default scenario, whose top vector score is 0.513), reducing
Avg~Hit@1 to 93.5\%.
At $\tau = 0.64$, Avg~Hit@1 drops sharply to 87.0\% as six
default-scenario queries are rejected, although
Miss-Empty-Rate reaches 100\%.
We recommend $\tau = 0.50$ as the operating point:
it ties with $\tau = 0.48$ on all accuracy metrics and is
the highest threshold that does not sacrifice recall.

\subsection{LongMemEval Evaluation}
\label{sec:longmemeval}

To evaluate MemX beyond the two custom scenarios (43 queries over
${\leq}1{,}014$ records), we apply it to the
LongMemEval benchmark~\citep{wang2024longmemeval}, which provides
500 annotated questions derived from 19{,}195 multi-turn
conversation sessions spanning four ability types:
\textit{information extraction} (156 queries),
\textit{knowledge update} (78),
\textit{multi-session reasoning} (133), and
\textit{temporal reasoning} (133).

\paragraph{Storage granularity.}
We convert the LongMemEval dataset into three MemX scenarios at
different storage granularities:
\begin{itemize}[nosep]
  \item \textbf{Session} — each conversation session becomes one
        memory (19{,}195 records).
  \item \textbf{Round} — each user--assistant turn pair becomes one
        memory (100{,}486 records).
  \item \textbf{Fact} — each atomic fact, extracted from sessions
        by an LLM (DeepSeek-Chat), becomes one memory
        (220{,}349 records).
\end{itemize}
This granularity axis is orthogonal to the pipeline-component
ablation in Section~\ref{sec:ablation}: it asks \emph{how should
raw conversations be chunked before indexing?}

\paragraph{Overall results.}
Table~\ref{tab:longmemeval_granularity} shows that finer
granularity consistently improves retrieval quality.
Fact-level storage doubles Hit@5 and MRR relative to
session-level storage.

\begin{table}[t]
\centering
\caption{LongMemEval retrieval quality across three storage
granularities (500 queries, $\tau{=}0.50$, FTS5 keyword index).}
\label{tab:longmemeval_granularity}
\begin{tabular}{@{}lrrrrr@{}}
\toprule
\textbf{Granularity}
  & \textbf{Records}
  & \textbf{Hit@1}
  & \textbf{Hit@5}
  & \textbf{MRR}
  & \textbf{TC@5} \\
\midrule
Session & 19{,}195  & 14.8\% & 24.6\% & 0.183 & 19.5\% \\
Round   & 100{,}486 & 17.2\% & 27.0\% & 0.207 & 21.0\% \\
Fact    & 220{,}349 & \textbf{29.8\%} & \textbf{51.6\%}
        & \textbf{0.380} & \textbf{41.5\%} \\
\bottomrule
\end{tabular}
\end{table}

\paragraph{Per-ability-type breakdown.}
Table~\ref{tab:longmemeval_ability} reveals that the improvement
from fact-level storage is not uniform across ability types.

\begin{table}[t]
\centering
\small
\caption{Hit@5 and MRR by LongMemEval ability type across three
storage granularities.}
\label{tab:longmemeval_ability}
\begin{tabular}{@{}lrcccccc@{}}
\toprule
& & \multicolumn{2}{c}{\textbf{Session}}
& \multicolumn{2}{c}{\textbf{Round}}
& \multicolumn{2}{c}{\textbf{Fact}} \\
\cmidrule(lr){3-4}\cmidrule(lr){5-6}\cmidrule(lr){7-8}
\textbf{Ability type} & $n$
  & Hit@5 & MRR
  & Hit@5 & MRR
  & Hit@5 & MRR \\
\midrule
Information extraction & 156
  & 36.5 & .298  & 39.7 & .324  & \textbf{55.8} & \textbf{.379} \\
Knowledge update & 78
  & 30.8 & .218  & 44.9 & .360  & \textbf{75.6} & \textbf{.624} \\
Multi-session reasoning & 133
  & 18.0 & .117  & 15.0 & .110  & \textbf{43.6} & \textbf{.323} \\
Temporal reasoning & 133
  & 13.5 & .093  & 13.5 & .078  & \textbf{40.6} & \textbf{.296} \\
\bottomrule
\end{tabular}
\end{table}

Three findings emerge.
First, \textbf{knowledge update} benefits most from fact-level
storage (Hit@5: $30.8\% \to 75.6\%$, $+44.8$\,pp), because
LLM-extracted facts explicitly surface the updated information
that a full conversation buries in dialogue context.
Second, \textbf{temporal reasoning} and
\textbf{multi-session reasoning} show the largest absolute gaps
from custom-scenario performance (${\leq}43.6\%$ Hit@5 at best),
confirming that these abilities require mechanisms beyond
single-query vector recall—e.g., temporal indexing or
cross-session linking—that MemX does not yet implement.
Third, moving from session to round granularity yields modest
gains ($+2.4$\,pp Hit@5 overall), while moving from round to
fact produces a large jump ($+24.6$\,pp), suggesting that
\emph{semantic density per record} is the primary driver of
retrieval quality at scale.

\paragraph{Pipeline baseline comparison.}
To isolate the contribution of pipeline components at this scale,
we compare a \textbf{vector-only baseline}
(no keyword search, no rejection, no deduplication) against the
\textbf{full pipeline} on the fact-granularity dataset.

\begin{table}[t]
\centering
\caption{Vector-only baseline vs.\ full pipeline on
LongMemEval fact granularity (220{,}349 records, 500 queries).}
\label{tab:baseline}
\begin{tabular}{@{}lrrrrr@{}}
\toprule
\textbf{Config}
  & \textbf{Hit@1}
  & \textbf{Hit@5}
  & \textbf{MRR}
  & \textbf{TC@5}
  & \textbf{Avg ms} \\
\midrule
V (vector-only) & \textbf{30.2\%} & \textbf{56.6\%}
  & \textbf{0.406} & \textbf{44.2\%} & 89.6 \\
Full pipeline    & 29.8\% & 51.6\%
  & 0.380 & 41.5\% & 88.2 \\
\bottomrule
\end{tabular}
\end{table}

Counter-intuitively, the full pipeline \emph{under}performs
the vector-only baseline by $5.0$\,pp on Hit@5
(Table~\ref{tab:baseline}).
The cause is tag-signature deduplication: in the fact-granularity
dataset, all memories are typed \texttt{semantic} with no tags,
producing an identical tag signature.
The deduplication rule therefore limits each query to at most
one result per signature, discarding co-relevant facts from the
same session that would otherwise occupy top-$k$ slots.
On the custom scenarios, where template-generated memories carry
explicit tags and $3{\times}$ scaling creates genuine duplicates,
deduplication improves Hit@3 by $+2.7$\,pp
(Table~\ref{tab:ablation}).
This contrast reveals that \textbf{deduplication is
data-dependent}: it helps when memories have structured tags and
repetitive content, but harms recall on tag-free atomic facts.
Section~\ref{sec:future} discusses adaptive deduplication
strategies that could address this.

\subsection{Latency Analysis}
\label{sec:latency}

\begin{table}[t]
\centering
\caption{Latency breakdown (milliseconds) across custom and
LongMemEval scenarios.
Custom scenarios use the remote embedding API;
LongMemEval scenarios use cached embeddings
(query\_embed ${\approx}\,0$\,ms).
Rows marked (LIKE) use \texttt{LIKE} substring search;
rows marked (FTS5) use FTS5 full-text indexing.
All runs use DiskANN vector indexing where supported.}
\label{tab:latency}
\begin{tabular}{@{}lrrrrr@{}}
\toprule
\textbf{Scenario}
  & \textbf{Records}
  & \textbf{Vector}
  & \textbf{Keyword}
  & \textbf{Total}
  & \textbf{P95} \\
\midrule
\multicolumn{6}{@{}l}{\textit{Custom scenarios (Section~\ref{sec:results})}} \\
Default        & 1{,}014  & 545   & 2.2  & 538   & 679   \\
High-Confusion & 600      & 389   & 1.6  & 398   & 460   \\
\midrule
\multicolumn{6}{@{}l}{\textit{LongMemEval (Section~\ref{sec:longmemeval})}} \\
Session (LIKE) & 19{,}195  & 98.6  & 286.7 & 385   & 1{,}418 \\
Session (FTS5) & 19{,}195  & 15.8  & 1.6  & 17.3  & 20    \\
Round (LIKE)   & 100{,}486 & 177.8 & 3{,}305 & 3{,}483 & 6{,}919 \\
Round (FTS5)   & 100{,}486 & 31.7  & 2.9  & 34.6  & 60    \\
Fact  (FTS5)   & 220{,}349 & 87.1  & 1.1  & 88.2  & 134   \\
\bottomrule
\end{tabular}
\end{table}

Table~\ref{tab:latency} reveals a scale-dependent latency picture.
On the custom scenarios (${\leq}1{,}014$ records), keyword search
is negligible ($<$3\,ms) and latency is dominated by the remote
embedding API call.
On LongMemEval at 100k+ records, however, the naive
\texttt{LIKE} keyword search becomes the dominant bottleneck:
at 100{,}486 records it consumes 3{,}305\,ms per query (95\% of
total latency), making it 1{,}500$\times$ slower than at 1k records.

Replacing \texttt{LIKE} with an FTS5 full-text index eliminates
this bottleneck entirely.
On the same 100{,}486-record dataset, FTS5 keyword search
takes 2.9\,ms—a \textbf{1{,}100$\times$} speedup—reducing total
search latency from 3{,}483\,ms to 34.6\,ms.
At 220{,}349 records (fact granularity), FTS5 keyword search
remains at 1.1\,ms and total latency is 88\,ms.

\begin{figure}[t]
  \centering
  \includegraphics[width=0.95\linewidth]{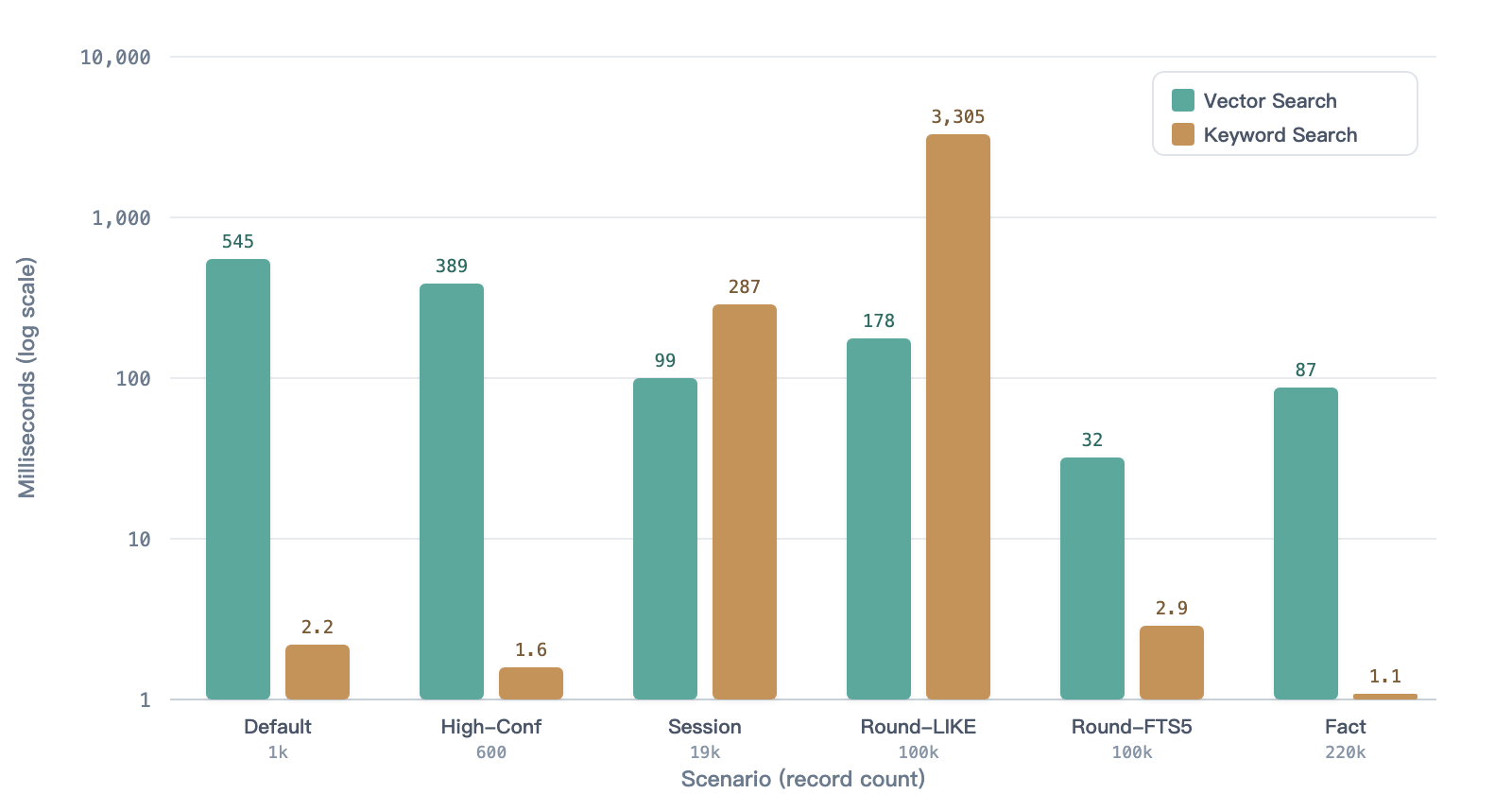}
  \caption{%
    Vector vs.\ keyword search latency across custom and
    LongMemEval scenarios (log scale).
    At 100k records, \texttt{LIKE}-based keyword search dominates
    total latency (3{,}305\,ms); FTS5 indexing reduces it to
    2.9\,ms ($1{,}100{\times}$ speedup).%
  }
  \label{fig:latency}
\end{figure}

Vector search latency also benefits from the DiskANN index
(Section~\ref{sec:arch}): at 100k records, DiskANN achieves
31.7\,ms compared to the brute-force $O(n)$ path that would
scale linearly with dataset size.
Figure~\ref{fig:latency} visualizes the contrast between
vector and keyword latency across all scenarios.

\subsection{Ablation Study}
\label{sec:ablation}

To quantify the contribution of each pipeline component, we
evaluate four cumulative configurations across both scenarios:
\textbf{V}~(vector search only),
\textbf{V+K}~(add keyword search and RRF fusion),
\textbf{V+K+Rej}~(add low-confidence rejection rule), and
\textbf{Full}~(add tag-signature deduplication).
All other parameters ($\tau{=}0.50$, four-factor weights,
$3{\times}$ scaling) are held constant.

\begin{table}[t]
\centering
\small
\caption{Ablation study: cumulative effect of pipeline components
averaged over both scenarios.
95\% Wilson confidence intervals are shown for binary metrics.
$n_\text{rel}$ and $n_\text{miss}$ denote the number of relevant
and miss queries respectively.}
\label{tab:ablation}
\begin{tabular}{@{}lccccc@{}}
\toprule
\textbf{Config}
  & \textbf{Hit@1}
  & \textbf{Hit@3}
  & \textbf{TC@$k$}
  & \textbf{Miss-Empty}
  & \textbf{Miss-Strict} \\
\midrule
V
  & 94.6 [82, 99]
  & 94.6 [82, 99]
  & 81.0
  & 0.0 [0, 39]
  & 83.3 [44, 97] \\
V+K
  & 94.6 [82, 99]
  & 94.6 [82, 99]
  & 81.0
  & 0.0 [0, 39]
  & 83.3 [44, 97] \\
V+K+Rej
  & 94.6 [82, 99]
  & 94.6 [82, 99]
  & 81.0
  & 66.7 [30, 90]
  & 83.3 [44, 97] \\
Full
  & 94.6 [82, 99]
  & \textbf{97.3} [86, 100]
  & \textbf{83.2}
  & 66.7 [30, 90]
  & 83.3 [44, 97] \\
\bottomrule
\multicolumn{6}{@{}l}{\scriptsize
  All values in \%.
  Pooled over $n_\text{rel}{=}37$ relevant and
  $n_\text{miss}{=}6$ miss queries.
  Wilson 95\% CI in brackets.}
\end{tabular}
\end{table}

Three observations emerge from Table~\ref{tab:ablation}.
First, adding keyword search and RRF fusion
(\textbf{V}$\to$\textbf{V+K}) produces no change in any metric.
This confirms that with the current query distribution, semantic
similarity alone is sufficient for ranking; keyword overlap adds
no discriminative power beyond what vector search already provides.

Second, the rejection rule
(\textbf{V+K}$\to$\textbf{V+K+Rej}) is the sole contributor to
Miss-Empty-Rate, raising it from 0\% to 66.7\%.
Without the rejection rule, the system returns results for every
query including those with no correct answer.
Hit@$k$ and TC@$k$ are unaffected, confirming that the rule
never rejects valid queries at $\tau{=}0.50$.

Third, deduplication
(\textbf{V+K+Rej}$\to$\textbf{Full}) lifts Hit@3 from 94.6\%
to 97.3\% and TC@$k$ from 81.0\% to 83.2\%
on the custom scenarios.
However, on LongMemEval's fact-granularity dataset the same
mechanism \emph{reduces} Hit@5 by $5.0$\,pp
(Section~\ref{sec:longmemeval}, Table~\ref{tab:baseline}).
This divergence arises because the custom scenarios use
structured tags that produce meaningful tag signatures, while
the fact dataset stores tag-free atomic statements that all
share an identical signature.
Deduplication is therefore \textbf{data-dependent}: beneficial
for tagged, template-generated data but harmful for
homogeneously-typed atomic facts.

\paragraph{Granularity ablation.}
A second, orthogonal ablation axis is explored in
Section~\ref{sec:longmemeval}: storage granularity
(session $\to$ round $\to$ fact).
Fact-level storage doubles Hit@5 and MRR relative to
session-level storage (Table~\ref{tab:longmemeval_granularity}),
an effect far larger than any pipeline-component change
observed above.
This suggests that for large-scale memory stores, investment in
upstream chunking and fact extraction yields greater returns
than additional retrieval pipeline stages.

\subsection{Rejection Rule Design Space}
\label{sec:rejection_space}

To justify the choice of rejection rule used throughout this work,
we define five candidate rules (Table~\ref{tab:rules}) and evaluate them against the
actual per-query data from both benchmark scenarios.
Let $v_{\max}$ denote the maximum vector similarity for a query
and $\mathit{kw}$ a Boolean indicating whether the keyword recall
set is non-empty.

\begin{table}[t]
\centering
\small
\caption{Five candidate rejection rules.
$v_{\max}$ is the top vector similarity score,
$\mathit{kw}$ indicates non-empty keyword recall,
and $\tau$ is the rejection threshold.}
\label{tab:rules}
\begin{tabular}{@{}clp{5.6cm}@{}}
\toprule
\textbf{Rule} & \textbf{Condition to reject} & \textbf{Intuition} \\
\midrule
R1 & $\lnot\mathit{kw} \;\land\; v_{\max} < \tau$ &
  Current system: reject only when \emph{both} signals are weak. \\
R2 & $v_{\max} < \tau$ &
  Vector-only gate; ignores keyword signal. \\
R3 & $\lnot\mathit{kw}$ &
  Keyword-required gate; ignores vector signal. \\
R4 & $\lnot\mathit{kw} \;\lor\; v_{\max} < \tau$ &
  Strict dual gate: reject if \emph{either} signal is weak. \\
R5 & $\lnot\mathit{kw} \;\land\; v_{\max} < 0.55$ &
  Same as R1 but with a raised threshold ($\tau{=}0.55$). \\
\bottomrule
\end{tabular}
\end{table}

We apply each rule to the original 24-query subset
(21 relevant $+$ 3 miss) for which per-query vector scores
were recorded.
Table~\ref{tab:rule_sim} summarizes the outcomes as
false negatives (valid queries incorrectly rejected) and
false positives (miss queries incorrectly accepted).
Four representative queries illustrate the key distinctions:
\texttt{job\_lookup} ($v_{\max}{=}0.504$, no~kw),
\texttt{incident\_keyword} ($v_{\max}{=}0.490$, kw),
\texttt{hard\_miss} ($v_{\max}{=}0.453$, no~kw),
and \texttt{public\_cloud\_region\_miss}
($v_{\max}{=}0.621$, no~kw).

\begin{table}[t]
\centering
\small
\caption{Rejection rule simulation on four representative queries.
FN = false negatives (valid queries rejected);
FP = false positives (miss queries accepted).
Counts are from the original 24-query subset where per-query
vector scores were recorded.}
\label{tab:rule_sim}
\begin{tabular}{@{}lccccl@{}}
\toprule
& \textbf{R1} & \textbf{R2} & \textbf{R3} & \textbf{R4} & \textbf{R5} \\
\midrule
\texttt{job\_lookup} (0.504, $\lnot$kw)
  & \checkmark & \checkmark & \texttimes & \texttimes & \texttimes \\
\texttt{incident\_kw} (0.490, kw)
  & \checkmark & \texttimes & \checkmark & \texttimes & \checkmark \\
\texttt{hard\_miss} (0.453, $\lnot$kw)
  & \texttimes & \texttimes & \texttimes & \texttimes & \texttimes \\
\texttt{cloud\_region\_miss} (0.621, $\lnot$kw)
  & \checkmark$^*$ & \checkmark$^*$ & \texttimes & \texttimes & \checkmark$^*$ \\
\midrule
FN (of 21 valid)  & \textbf{0} & 1 & 18 & 19 & 1 \\
FP (of 3 miss)    & 1 & 1 & 1 & 0 & 1 \\
\bottomrule
\multicolumn{6}{@{}l}{\scriptsize
  \checkmark\ = correctly handled;\quad
  \texttimes\ = incorrectly rejected/accepted;\quad
  $^*$ = false positive (miss query accepted).}
\end{tabular}
\end{table}

Three findings emerge from this analysis.

First, \textbf{R1 is the only rule with zero false negatives}.
It preserves all 21 valid queries because its conjunctive structure
($\lnot\mathit{kw} \land v_{\max}{<}\tau$) requires both signals to
be absent before rejecting.
All other rules sacrifice at least one valid query:
R2 rejects \texttt{incident\_keyword} ($v_{\max}{=}0.490 < \tau$
despite a keyword hit);
R3 and R4 reject 18 and 19 valid queries respectively because
only 3 of the 24 queries produce keyword hits;
R5 rejects \texttt{job\_lookup} ($v_{\max}{=}0.504 < 0.55$).

Second, \textbf{R3 and R4 are unusable in practice}.
Because keyword recall relies on token overlap between query
and stored memories (whether via FTS5 or substring matching),
only 3 of 24 queries produce non-empty keyword sets.
Any rule that requires a keyword hit as a necessary condition for
acceptance will reject almost all valid queries.

Third, \textbf{the fundamental limit is the score gap}
between the hardest valid query (\texttt{job\_lookup}, $v_{\max}{=}0.504$)
and the easiest false recall
(\texttt{public\_cloud\_region\_miss}, $v_{\max}{=}0.621$).
No single-threshold rule over $v_{\max}$ can separate these two
queries: any $\tau$ that accepts \texttt{job\_lookup} must also
accept the false recall.
Improving miss suppression beyond the current level therefore
requires a different kind of signal, such as cross-encoder
re-scoring, which can model fine-grained query--document
interaction rather than relying on a single similarity score.

\section{Discussion}
\label{sec:discussion}

\subsection{Failure Cases and System Boundaries}
\label{sec:failures}

\paragraph{Multi-topic coverage gap.}
In both scenarios, queries that require two topics consistently
retrieve only the most semantically prominent one.
For example, the query \textit{``Which metric should matter
besides budget when recommending a tool?''} retrieves
\texttt{budget\_threshold} but not \texttt{search\_expectation}.
The current pipeline is optimized for single-topic
``needle-in-a-haystack'' retrieval and lacks mechanisms for
multi-facet query decomposition.

\paragraph{High-confusion false recall.}
The \texttt{public\_cloud\_region\_miss} query shares substantial
vocabulary with \texttt{local\_storage\_policy} (both discuss
deployment, data location, and cloud infrastructure).
The resulting vector similarity (0.621) exceeds $\tau$, producing
a false recall.
The rejection rule, while effective for topically distant
miss queries, can be defeated by carefully constructed
adversarial overlaps.
The systematic analysis in Section~\ref{sec:rejection_space} confirms
that no single-threshold rule can separate this false recall from
valid queries.

\paragraph{Graph structure not yet in the search path.}
The \texttt{memory\_links} table and its seven relation types
provide the scaffolding for graph-enhanced retrieval.
However, the current search pipeline does not perform multi-hop
traversal or graph-aware re-ranking, so the graph cannot yet
resolve contradictions or surface causally linked memories.

\paragraph{Temporal and multi-session reasoning.}
The LongMemEval evaluation (Section~\ref{sec:longmemeval})
reveals that temporal reasoning (${\leq}40.6\%$ Hit@5 at
fact granularity) and multi-session reasoning
(${\leq}43.6\%$) are the weakest ability types.
These queries require capabilities that single-query vector
recall cannot provide: temporal ordering of events across
sessions, and synthesis of information scattered across
multiple conversations.
Addressing these gaps likely requires temporal indexing
(e.g., date-aware retrieval) and cross-session linking
mechanisms.

\paragraph{No task-level attribution.}
The system tracks which memories are returned by search, but
does not record whether a returned memory was actually
\emph{used} by the downstream agent or whether it improved
task completion.
Without this closed-loop signal, the system cannot
automatically learn which memories are most valuable.

\subsection{Impact of Access/Retrieval Separation}
\label{sec:separation_impact}

Section~\ref{sec:tracking} described the design rationale for
separating access and retrieval counters.
Because the benchmark creates all memories fresh with zero counters,
it cannot directly exercise this design choice: recency and frequency
factors contribute equally (and minimally) regardless of tracking
strategy.
To illustrate the practical impact, we construct a worked example
with three hypothetical memory profiles exhibiting different
access/retrieval patterns.

\begin{table}[t]
\centering
\small
\caption{Three memory profiles with contrasting access and
retrieval histories.
All three share $f_{\text{sem}} = 0.70$ and $f_{\text{imp}} = 0.80$.}
\label{tab:profiles}
\begin{tabular}{@{}llrrll@{}}
\toprule
& \textbf{Profile}
& \textbf{Access}
& \textbf{Retrieval}
& \textbf{Last accessed}
& \textbf{Last retrieved} \\
\midrule
A & Admin-heavy  & 50 & 2  & 1\,d ago  & 15\,d ago \\
B & Search-heavy & 3  & 25 & 30\,d ago & 1\,d ago  \\
C & Stale        & 40 & 40 & 60\,d ago & 60\,d ago \\
\bottomrule
\end{tabular}
\end{table}

We compute the composite score (Equation~\ref{eq:rerank}) for each
profile under two strategies.
\textbf{Strategy~1} (retrieval-preferred, the current design) uses
\texttt{retrieval\_count} for frequency and
\texttt{last\_re\-trieved\_at} for recency.
\textbf{Strategy~2} (access-based) uses
\texttt{access\_count} for frequency and
\texttt{last\_ac\-cessed\_at} for recency.
Both strategies use the same factor formulas defined in
Section~\ref{sec:rerank}: recency
$f_{\text{rec}} = 2^{-d/h}$ with $h{=}30$ days
(Equation~\ref{eq:recency}) and frequency
$f_{\text{freq}} = \min(1,\,\ln(c{+}1)/10)$
(Equation~\ref{eq:frequency}),
differing only in which timestamp and counter they read.
Semantic and importance factors are held constant at
$f_{\text{sem}} = 0.70$ and $f_{\text{imp}} = 0.80$.

\begin{table}[t]
\centering
\small
\caption{Composite scores under two tracking strategies.
Weights: $\alpha_s{=}0.45$, $\alpha_r{=}0.25$,
$\alpha_f{=}0.05$, $\alpha_i{=}0.10$.
Ranking reversal between A and B is highlighted.}
\label{tab:separation}
\begin{tabular}{@{}lcccccc@{}}
\toprule
& \multicolumn{3}{c}{\textbf{Strategy 1 (retrieval)}}
& \multicolumn{3}{c}{\textbf{Strategy 2 (access)}} \\
\cmidrule(lr){2-4}\cmidrule(lr){5-7}
& $f_{\text{rec}}$ & $f_{\text{freq}}$ & Score
& $f_{\text{rec}}$ & $f_{\text{freq}}$ & Score \\
\midrule
A & 0.707 & 0.110 & 0.577
  & 0.977 & 0.393 & 0.659 \\
B & 0.977 & 0.326 & \textbf{0.656}
  & 0.500 & 0.139 & 0.527 \\
C & 0.250 & 0.371 & 0.476
  & 0.250 & 0.371 & 0.476 \\
\midrule
\multicolumn{1}{@{}l}{Ranking}
  & \multicolumn{3}{c}{B $>$ A $>$ C}
  & \multicolumn{3}{c}{A $>$ B $>$ C} \\
\bottomrule
\end{tabular}
\end{table}

The ranking reversal between profiles~A and~B is driven almost
entirely by the recency factor ($\alpha_r = 0.25$), not by
frequency ($\alpha_f = 0.05$).
Under Strategy~1, profile~B ranks first because it was
\emph{retrieved} one day ago; under Strategy~2, profile~A ranks
first because it was \emph{accessed} one day ago.
Strategy~2 therefore promotes memories that are frequently viewed
(e.g., for administrative purposes) but rarely useful in search,
while demoting memories that are genuinely search-relevant.
This confirms the design motivation in Section~\ref{sec:tracking}:
retrieval-based tracking better reflects a memory's utility in the
search pipeline.

\subsection{Threats to Validity}
\label{sec:threats}

\begin{enumerate}[leftmargin=*,nosep]
  \item \textbf{Template-driven data.}
        Benchmark records are generated from templates with
        repetition-based scaling.
        While more structured than ad-hoc examples, they do not
        capture the distributional complexity of real long-term
        interaction logs.
  \item \textbf{Single embedding configuration.}
        All results are obtained with Qwen3-Embedding-0.6B at
        1{,}024 dimensions.
        The system accepts any OpenAI-compatible embedding model
        via environment variables, but the optimal rejection
        threshold~$\tau$ must be recalibrated for each model,
        as cosine-similarity distributions vary across
        embedding spaces.
  \item \textbf{Intrinsic vs.\ end-to-end latency.}
        Reported latencies bypass HTTP routing and full request
        handling; production latency will be higher.
  \item \textbf{No human evaluation.}
        The current evaluation measures retrieval quality, not
        downstream task completion.
        Whether improved retrieval translates to better agent
        output requires task-level human assessment.
  \item \textbf{Scenario diversity.}
        The custom evaluation uses 43 queries across two
        scenarios; the addition of LongMemEval (500 queries,
        four ability types) substantially broadens coverage,
        but all LongMemEval conversations are in English while
        the custom scenarios are in Chinese.
        Cross-lingual generalization remains untested.
\end{enumerate}

\subsection{Future Directions}
\label{sec:future}

Five directions are most promising for extending the current baseline:
\begin{enumerate}[leftmargin=*,nosep]
  \item \textbf{Temporal and cross-session reasoning.}
        The LongMemEval evaluation identifies temporal reasoning
        and multi-session reasoning as the weakest ability types.
        Date-aware retrieval (e.g., temporal indexing of memory
        creation times) and cross-session linking could
        substantially improve these categories.
  \item \textbf{Stronger rejection rules.}
        Cross-encoder re-scoring or lightweight classifier-based
        rejection could improve miss-query handling under high
        semantic overlap.
  \item \textbf{Adaptive deduplication.}
        The current tag-signature deduplication helps on
        template-generated data but harms recall on tag-free
        atomic facts (Section~\ref{sec:longmemeval}).
        Content-similarity--based deduplication (e.g.,
        collapsing results only when cosine similarity exceeds
        a threshold) would adapt to both regimes without
        requiring explicit tags.
  \item \textbf{Multi-topic coverage.}
        Query decomposition (splitting a compound query into
        sub-queries) or diversity-aware re-ranking could
        improve Coverage@$k$ on multi-facet queries.
  \item \textbf{Task-level attribution.}
        Recording whether a retrieved memory was used by the
        agent and whether it improved the final output would
        enable credit assignment and a closed-loop memory
        lifecycle.
\end{enumerate}

\section{Conclusion}
\label{sec:conclusion}

We have presented MemX, a local-first long-term memory system
for AI assistants, featuring a hybrid retrieval pipeline with
RRF fusion, four-factor re-ranking, and a low-confidence
rejection rule.

On two custom Chinese-language scenarios (43 queries),
the pipeline achieves $\text{Hit@1}{=}91$--$100\%$ with
conservative miss-query suppression, validating its
component design.
On the LongMemEval benchmark (500 queries, up to 220k records),
fact-level storage reaches $\text{Hit@5}{=}51.6\%$ and
$\text{MRR}{=}0.380$—doubling session-level
performance—while temporal and multi-session reasoning
remain at ${\leq}43.6\%$ Hit@5, defining clear improvement
targets.
A storage-granularity study shows that semantic density per
record is the primary driver of retrieval quality at scale,
with knowledge-update queries benefiting most
($\text{Hit@5}$: $30.8\% \to 75.6\%$).

On the systems side, replacing \texttt{LIKE}-based keyword
search with FTS5 full-text indexing yields a
$1{,}100{\times}$ latency reduction at 100k records,
keeping end-to-end search under 90\,ms at 220k records.

Two supplementary analyses strengthen these findings:
a rejection rule design space study
(Section~\ref{sec:rejection_space}) shows that the current
conjunctive rule is the only candidate with zero false negatives,
and an access/retrieval separation analysis
(Section~\ref{sec:separation_impact}) demonstrates that
retrieval-based tracking prevents ranking distortion from
administrative access patterns.

Within the scope of a local, single-user AI assistant, MemX
provides a solid v1 baseline whose retrieval pipeline,
explicit tracking separation, and reproducible benchmark
framework can be extended with temporal indexing, cross-session
linking, and task-level attribution without architectural
redesign.
Source code and reproduction scripts are available at
\url{https://github.com/memxlab/memx}.

\bibliography{references}

@article{chhikara2025mem0,
  title   = {Mem0: Building Production-Ready AI Agents with Scalable Long-Term Memory},
  author  = {Chhikara, Prateek and Khindri, Vansh and Udasi, Deshraj and Thakkar, Dev},
  journal = {arXiv preprint arXiv:2504.19413},
  year    = {2025},
}

@inproceedings{packer2024memgpt,
  title     = {{MemGPT}: Towards {LLM}s as Operating Systems},
  author    = {Packer, Charles and Wooders, Sarah and Lin, Kevin and Fang, Vivian and Patil, Shishir G. and Stoica, Ion and Gonzalez, Joseph E.},
  booktitle = {Proceedings of the International Conference on Learning Representations (ICLR)},
  year      = {2024},
}

@inproceedings{park2023generative,
  title     = {Generative Agents: Interactive Simulacra of Human Behavior},
  author    = {Park, Joon Sung and O'Brien, Joseph C. and Cai, Carrie J. and Morris, Meredith Ringel and Liang, Percy and Bernstein, Michael S.},
  booktitle = {Proceedings of the 36th Annual ACM Symposium on User Interface Software and Technology (UIST)},
  year      = {2023},
}

@article{zhang2024survey,
  title   = {A Survey on the Memory Mechanism of Large Language Model Based Agents},
  author  = {Zhang, Zeyu and Zhang, Xiaohe Bo and Jiang, Chen and Liu, Bin and Fan, Zhenyu and Guo, Fei and Gong, Mingyu},
  journal = {arXiv preprint arXiv:2404.13501},
  year    = {2024},
}

@article{modarressi2024memllm,
  title   = {{MemLLM}: Finetuning {LLM}s to Use An Explicit Read-Write Memory},
  author  = {Modarressi, Ali and Imani, Ayyoob and Fayyaz, Mohsen and Sch{\"u}tze, Hinrich},
  journal = {arXiv preprint arXiv:2404.11672},
  year    = {2024},
}

@inproceedings{lewis2020rag,
  title     = {Retrieval-Augmented Generation for Knowledge-Intensive {NLP} Tasks},
  author    = {Lewis, Patrick and Perez, Ethan and Piktus, Aleksandra and Petroni, Fabio and Karpukhin, Vladimir and Goyal, Naman and K{\"u}ttler, Heinrich and Lewis, Mike and Yih, Wen-tau and Rockt{\"a}schel, Tim and Riedel, Sebastian and Kiela, Douwe},
  booktitle = {Advances in Neural Information Processing Systems (NeurIPS)},
  year      = {2020},
}

@article{gao2024rag_survey,
  title   = {Retrieval-Augmented Generation for Large Language Models: A Survey},
  author  = {Gao, Yunfan and Xiong, Yun and Gao, Xinyu and Jia, Kangxiang and Pan, Jinliu and Bi, Yuxi and Dai, Yi and Sun, Jiawei and Wang, Meng and Wang, Haofen},
  journal = {arXiv preprint arXiv:2312.10997},
  year    = {2024},
}

@article{johnson2019faiss,
  title   = {Billion-Scale Similarity Search with {GPU}s},
  author  = {Johnson, Jeff and Douze, Matthijs and J{\'e}gou, Herv{\'e}},
  journal = {IEEE Transactions on Big Data},
  volume  = {7},
  number  = {3},
  pages   = {535--547},
  year    = {2021},
}

@article{malkov2020hnsw,
  title   = {Efficient and Robust Approximate Nearest Neighbor Search Using Hierarchical Navigable Small World Graphs},
  author  = {Malkov, Yu A. and Yashunin, D. A.},
  journal = {IEEE Transactions on Pattern Analysis and Machine Intelligence},
  volume  = {42},
  number  = {4},
  pages   = {824--836},
  year    = {2020},
}

@inproceedings{cormack2009rrf,
  title     = {Reciprocal Rank Fusion Outperforms {Condorcet} and Individual Rank Learning Methods},
  author    = {Cormack, Gordon V. and Clarke, Charles L. A. and Buettcher, Stefan},
  booktitle = {Proceedings of the 32nd International ACM SIGIR Conference on Research and Development in Information Retrieval},
  pages     = {758--759},
  year      = {2009},
}

@inproceedings{chen2024bge_m3,
  title     = {{BGE M3}-Embedding: Multi-Lingual, Multi-Functionality, Multi-Granularity Text Embeddings Through Self-Knowledge Distillation},
  author    = {Chen, Jianlv and Xiao, Shitao and Zhang, Peitian and Luo, Kun and Lian, Defu and Liu, Zheng},
  booktitle = {Proceedings of the 62nd Annual Meeting of the Association for Computational Linguistics (ACL)},
  year      = {2024},
}

@inproceedings{lin2021bm25,
  title     = {Pyserini: A Python Toolkit for Reproducible Information Retrieval Research with Sparse and Dense Representations},
  author    = {Lin, Jimmy and Ma, Xueguang and Lin, Sheng-Chieh and Yang, Jheng-Hong and Pradeep, Ronak and Nogueira, Rodrigo},
  booktitle = {Proceedings of the 44th International ACM SIGIR Conference on Research and Development in Information Retrieval},
  pages     = {2356--2362},
  year      = {2021},
}

@article{qwen3embedding2025,
  title   = {Qwen3 Embedding and Reranker: Towards Unified Versatile Embedding and Reranking},
  author  = {{Qwen Team}},
  journal = {Qwen Technical Blog},
  year    = {2025},
  note    = {\url{https://qwenlm.github.io/blog/qwen3-embedding/}},
}

@article{edge2024graphrag,
  title   = {{GraphRAG}: Unlocking {LLM} Discovery on Narrative Private Data},
  author  = {Edge, Darren and Trinh, Ha and Cheng, Newman and Bradley, Joshua and Chao, Alex and Mody, Apurva and Truitt, Steven and Larson, Jonathan},
  journal = {arXiv preprint arXiv:2404.16130},
  year    = {2024},
}

@article{guo2024knowledgegraph_survey,
  title   = {A Survey on Knowledge Graph-Enhanced Large Language Models},
  author  = {Guo, Jiajun and others},
  journal = {arXiv preprint arXiv:2404.14741},
  year    = {2024},
}

@article{kleppmann2019localfirst,
  title   = {Local-First Software: You Own Your Data, in spite of the Cloud},
  author  = {Kleppmann, Martin and Wiggins, Adam and van Hardenberg, Peter and McGranaghan, Mark},
  journal = {Proceedings of the ACM on Human-Computer Interaction},
  volume  = {3},
  number  = {CSCW},
  pages   = {1--24},
  year    = {2019},
}

@misc{hipp2000sqlite,
  title        = {{SQLite}},
  author       = {Hipp, D. Richard},
  year         = {2000},
  howpublished = {\url{https://www.sqlite.org/}},
}

@misc{libsql2023,
  title        = {{libSQL}: An Open Contribution Fork of {SQLite}},
  author       = {{Turso}},
  year         = {2023},
  howpublished = {\url{https://github.com/tursodatabase/libsql}},
}

@article{wilson1927,
  title   = {Probable Inference, the Law of Succession, and Statistical Inference},
  author  = {Wilson, Edwin B.},
  journal = {Journal of the American Statistical Association},
  volume  = {22},
  number  = {158},
  pages   = {209--212},
  year    = {1927},
}

@article{wang2024longmemeval,
  title   = {LongMemEval: Benchmarking Chat Assistants on Long-Term Interactive Memory},
  author  = {Wang, Di and others},
  journal = {arXiv preprint arXiv:2410.10813},
  year    = {2024},
}

\end{document}